\documentclass[aps,longbibliography,showpacs,twocolumn,superscriptaddress]{revtex4-1}
\usepackage{amsmath,amssymb,amsfonts,bm}
\usepackage{graphicx}
\usepackage{epstopdf}
\usepackage{dcolumn}
\usepackage{mathrsfs}
\usepackage{bbold}
\usepackage{dsfont}
\usepackage{float}
\usepackage[colorlinks=true,linkcolor=blue,citecolor=blue, urlcolor=blue,bookmarks=false]{hyperref}

\usepackage{tgtermes}
\usepackage{multirow}
\begin{document}

\title{Higher-order topological insulator in a dodecagonal quasicrystal}
\date{\today }
\author{Chun-Bo Hua}
\affiliation{Department of Physics, Hubei University, Wuhan 430062, China}
\author{Rui Chen}
\affiliation{Shenzhen Institute for Quantum Science and Engineering and Department of Physics, Southern University of Science and Technology, Shenzhen 518055, China}
\affiliation{School of Physics, Southeast University, Nanjing 211189, China}
\author{Bin Zhou}\email{binzhou@hubu.edu.cn}
\affiliation{Department of Physics, Hubei University, Wuhan 430062, China}
\author{Dong-Hui Xu}\email{donghuixu@hubu.edu.cn}
\affiliation{Department of Physics, Hubei University, Wuhan 430062, China}

\begin{abstract}
Higher-order topological insulators (HOTIs) are a newly discovered class of topological insulators which exhibit unconventional bulk-boundary correspondence. Very recently, the concept of HOTIs has been extended to aperiodic quasicrystalline systems, where the topological band theory fails to describe topological phases. More importantly, a novel HOTI phase protected by an eightfold rotational symmetry, not found in crystalline materials, was identified in the Ammann-Beenker tiling quasicrystals. Here we report the discovery of a quasicrystalline HOTI in a dodecagonal quasicrystal. The quasicrystalline HOTI supports twelve in-gap zero-energy modes symmetrically distributed at the corners of the quasicrystal dodecagon. These zero-energy corner modes are protected by a combination of a twelvefold rotational symmetry and mirror symmetry, as well as particle-hole symmetry, which has no crystalline counterpart.
\end{abstract}

\maketitle

\emph{\color{blue}Introduction.}---
Since the discovery of topological insulators, the exploration of various topological phases of matter has become a major goal of research in condensed matter physics \cite{Hasan2010RMP,Qi2011RMP,Bansil2016RMP,Haldane2017RMP,Wen2017RMP,Wolfle2018RPP}. According to  the three fundamental nonspatial symmetries containing particle-hole symmetry (PHS), time-reversal symmetry (TRS), and chiral symmetry, the fully
gapped free fermionic systems have been classified into the ten Altland-Zirnbauer symmetry classes \cite{AZ1997PRB,Schnyder2008PRB,Schnyder2009AIP,Ryu2010NJP,Chiu2016RMP}. Later, with the notion of ``topological crystalline insulators"~\cite{Fu2011PRL,Fu2015ARCMP} being put forward, the influence of space-group symmetries on topological phases has attracted extensive attention and enriches the topological classification of crystalline solids~\cite{Fu2011PRL,Fu2015ARCMP,PhysRevB.88.075142,PhysRevB.88.125129,PhysRevB.90.165114,lu2014inversion,
PhysRevB.93.195413,PhysRevB.96.195109,Slager_2012,PhysRevX.7.041069}. Recently, a new class of topological crystalline insulators, known as higher-order topological insulators~(HOTIs) \cite{Zhang2013PRL,Benalcazar2017Science,Langbehn2017PRL,Song2017PRL,Benalcazar2017PRB,
Schindler2018SA,Ezawa2018PRL,Ezawa2018PRL2,
Ezawa2018PRB2,Geier2018PRB,Khalaf2018PRB,Ezawa2018PRB1,Kunst2018PRB,
Ezawa2018PRB,Guido2018PRB,Franca2018PRB,You2018PRB,Mao2018PRB,Kooi2018PRB,
Luka2019PRX,Ahn2019PRX,Lee2019PRL,Tao2019PRL,Feng2019PRL,Fan2019PRL,Zhijun2019PRL,Pozo2019PRL,Sheng2019PRL,
PhysRevLett.124.036803,Varjas2019PRL,Roy2019PRB,Benalcazar2019PRB,Rodriguez2019PRB,Hwang2019PRB,Okugawa2019PRB,
PhysRevResearch.2.012067,PhysRevB.101.041404,PhysRevLett.121.096803,Wang2018PRL,
Shapourian2018PRB,Dwivedi2018PRB,Yuxuan2018PRB,Tao2018PRB,
Liu2019PRL,Yan2019PRL,Zhu2019PRL,Bultinck2019PRB,Yan2019PRB,Fulga2019PRB,PhysRevLett.125.097001,
Serra2018Nature,Peterson2018Nature,Xue2018NM,Ni2018NM,Schindler2018NP,Imhof2018NP,Noh2018NPh,
Zhang2019NP,Kempkes2019NM,Mittal2019NPh,Hassan2019NPh,Lee2020npjQM,PhysRevLett.123.196402,PhysRevLett.122.256402,Yue2019NP,
PhysRevLett.121.196801,PhysRevLett.122.126402,PhysRevB.92.085126}, was discovered. Like previously known topological crystalline insulators, HOTIs need the protection of space-group symmetries such as mirror, inversion, and rotational
symmetries, but exhibit unconventional bulk-boundary correspondence. For instance, a second-order topological insulator in two dimensions supports topological gapless boundary states at its zero-dimensional boundary corners, in contrast to the conventional two-dimensional (2D) first-order topological insulators which have one-dimensional (1D) gapless edge states.

Up to now, most of discovered topological phases exist in crystalline systems, which can be characterized by the topological band theory. Interestingly, the quasicrystalline systems, which lack the translational discrete symmetry and may possess forbidden rotational symmetries in crystals, have been found to host topological phases \cite{PhysRevX.6.011016,PhysRevLett.121.126401,PhysRevB.98.125130,PhysRevB.100.214109,PhysRevX.9.021054,
PhysRevB.100.085119,PhysRevB.91.085125,PhysRevB.94.205437,PhysRevB.100.115311,PhysRevB.101.115413,
PhysRevLett.124.036803,Varjas2019PRL,PhysRevB.101.041103,PhysRevLett.109.106402,Kraus2012PRL,Ganeshan2013PRL,Longhi2019PRL,
Fulga2016PRL,Yichen2018PRL,Poyhonen2018NC}. Even more striking is that a kind of exotic HOTI protected by an eightfold rotational symmetry
is proposed in the Ammann-Beenker (AB) tiling octagonal quasicrystal~\cite{PhysRevLett.124.036803, Varjas2019PRL}, which cannot be found in crystals because an eightfold rotational symmetry is incompatible with translational symmetry. Therefore, quasicrystalline systems with forbidden rotational symmetries offer a platform to explore novel topological phases beyond the present topological classification of crystalline materials.

\begin{figure}[t]
	\includegraphics[width=5cm]{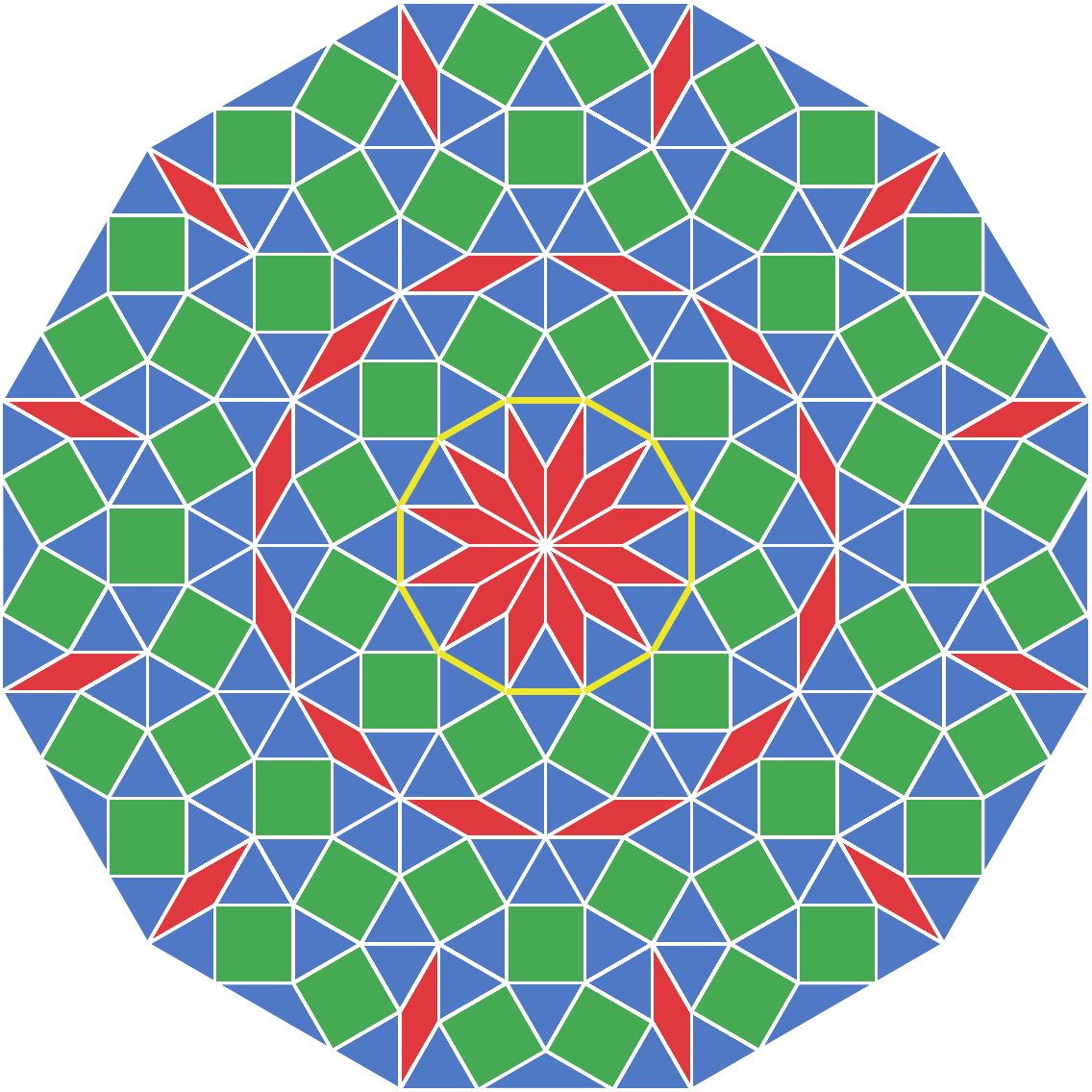} \caption{Schematic illustrations of the Stampfli-tiling quasicrystal dodecagon. The quasicrystal consists of three types of primitive tiles: square tiles (green), regular triangle tiles (blue), and rhombus tiles (red) with a small angle  $30^{\circ}$.}%
	\label{fig1}
\end{figure}
Another representative type of quasicrystal is the dodecagonal quasicrystal with twelvefold rotational symmetry. The first dodecagonal quasicrystalline lattice~(QL) was proposed by P. Stampfli \cite{stampfli1986dodecagonal}. Recently, the dodecagonal QL was realized in the twisted bilayer graphene rotated exactly ${30}^{\ensuremath{\circ}}$ ~\cite{Koren2016PRB,Ahn2018Science,Yao2018PNAS,PhysRevB.99.165430,Yu2019NPJCM}, which makes the dodecagonal quasicrystal more attractive. Meanwhile, Park $et\ al$.~\cite{PhysRevLett.123.216803} proposed that a crystalline HOTI protected by a mirror symmetry and a sixfold rotational symmetry can be realized in the twisted bilayer graphene with the twist angle $21.78^{\ensuremath{\circ}}$.

In this work, we investigate the 2D quasicrystalline HOTI phase in the Stampfli-tiling dodecagonal quasicrystal. The dodecagonal quasicrystal is tiled using squares, triangles, and rhombuses (see Fig.~\ref{fig1}). The construction process of dodecagonal quasicrystals through the inflation method is shown in Ref. \cite{SM} (see Fig. S1). We uncover a quasicrystalline HOTI protected by a combination of the twelvefold rotational symmetry $C_{12}$ and mirror symmetry $m_z$ in the Stampfli-tiling dodecagonal quasicrystal. To obtain the HOTI, we start with a first-order topological insulator model with TRS in the dodecagonal quasicrystal which supports counterpropagating 1D edge modes. Then, we introduce an additional mass term that breaks both TRS and $C_{12}$ but preserves the combined symmetry $C_{12}m_z$. 1D gapless edge modes are gapped out, and twelve ingap zero-energy modes emerge symmetrically at the boundary corners of a regular Stampfli-tiling quasicrystal dodecagon, which are the hallmark feature of the 2D quasicrystalline HOTI. These corner modes are robust against any symmetry-preserving perturbations.

\emph{\color{blue}Counterpropagating edge modes on the dodecagonal QL.}---We first consider a first-order topological insulator tight-binding model on the dodecagonal QL. The lattice sites are located on the vertices of the Stampfli-tiling as shown in Fig.~\ref{fig1}. The nearest-neighbor sites are connected by short diagonals of rhombuses, and the next-nearest-neighbor sites are connected by the sides of the three primitive tiles, etc. The model Hamiltonian in the Stampfli-tiling quasicrystal dodecagon is given by
\begin{align}
H_{0}=&-\sum_{j\not=k}\frac{u\left(r_{jk}\right)}{2}c_{j}^{\dag }\left[it_{1}\left(\sigma_{z}\tau_{x}\cos\phi_{jk}+\sigma_{0}\tau _{y}\sin\phi_{jk} \right) \right. \nonumber \\
\phantom{=;\;\;}
&\left. +t_{2}\sigma_{0}\tau_{z} \right]c_{k}+\sum_{j}\left(M+2t_{2}\right)c_{j}^{\dag }\sigma_{0}\tau_{z}c_{j},
\label{H0}
\end{align}
where the basis is $c_{j}^{\dag}=(c_{j\alpha\uparrow}^{\dag },c_{j\alpha\downarrow}^{\dag },c_{j\beta\uparrow}^{\dag },c_{j\beta\downarrow}^{\dag })$, $\alpha$ and $\beta$ are different orbital degree of freedom, $\uparrow$ and $\downarrow$ represent electron spin, $j$ and $k$ denote lattice sites running from $1$ to $N$, and $N$ is the total number of lattice sites. $\sigma_{0}$ and $\tau_{0}$ are two $2\times2$ identity matrices, $\sigma_{x,y,z}$ and $\tau_{x,y,z}$ are the Pauli matrices acting on the spin and orbital degrees of freedom, respectively. $M$ is the mass that determines the topological insulator phase, $t_{1}$ and $t_{2}$ are hopping parameters. $\phi_{jk}$ is the polar angle of bond connecting sites $j$ and $k$ with respect to the horizontal direction. $u\left( d_{jk}\right)=e^{-(d_{jk}-1)/\xi }$ is the spatial decay factor of hoppings with the decay length $\xi$, and $d_{jk}=\left\vert \mathbf{d}_{j}-\mathbf{d}_{k}\right\vert$ is the lattice site distance, where $\mathbf{d}_{j}$ and $\mathbf{d}_{k}$ are the coordinates of lattice sites. In subsequent calculations, the energy unit is set as $t_{2}$, the next-nearest-neighbor lattice distance is used as the length unit, and the spatial decay length $\xi$ is fixed as $1$.

Before we move to the energy spectrum of edge states in the topological insulator, we would like to discuss the symmetries of the Hamiltonian $H_0$. The Hamiltonian (\ref{H0}) satisfies
\begin{equation}
PH_{0}P^{-1}=-H_{0}, TH_{0}T^{-1}=H_{0}, SH_{0}S^{-1}=-H_{0}.
\end{equation}
Here $P$, $T$, $S$ are PHS, TRS and chiral symmetry operators, respectively, and they are expressed by
\begin{equation}
P=\sigma_{z}\tau_{x}\mathcal{I}K,  T=i \sigma_{y}\tau_{0}\mathcal{I}K,  S=PT,
\end{equation}
where $K$ is the complex conjugate operator, and $\mathcal{I}$ is the $N\times N$ identity matrix.
Therefore, $H_0$ possesses PHS, TRS, and chiral symmetry, and belongs to the symmetry class DIII~ \cite{AZ1997PRB,Schnyder2008PRB,Schnyder2009AIP,Ryu2010NJP,Chiu2016RMP}. In addition, $H_0$ has a mirror symmetry about the $x$-$y$ plane $m_z$, and satisfies $[H_{0},m_{z}]=0$ with $m_{z}= \sigma_{z}\tau_{0}\mathcal{I}$. Simultaneously, for the Stampfli-tiling quasicrystal dodecagon, the Hamiltonian $H_0$ also satisfies $[H_{0},C_{12}]=0$, thus it has twelvefold rotational symmetry. Here the twelvefold rotational symmetry operator is $C_{12}=e^{-i\frac{\pi}{12}\sigma_{z}\tau_{z}}\mathcal{R}_{12}$, where $\mathcal{R}_{12}$ is an orthogonal matrix permuting the sites of the QL to rotate the whole system by an angle of $\pi/6$. We give more details of symmetry analysis for the Hamiltonian (\ref{H0}) in Ref. \cite{SM} (see Tab. S1).

\begin{figure}[t]
	\includegraphics[width=8cm]{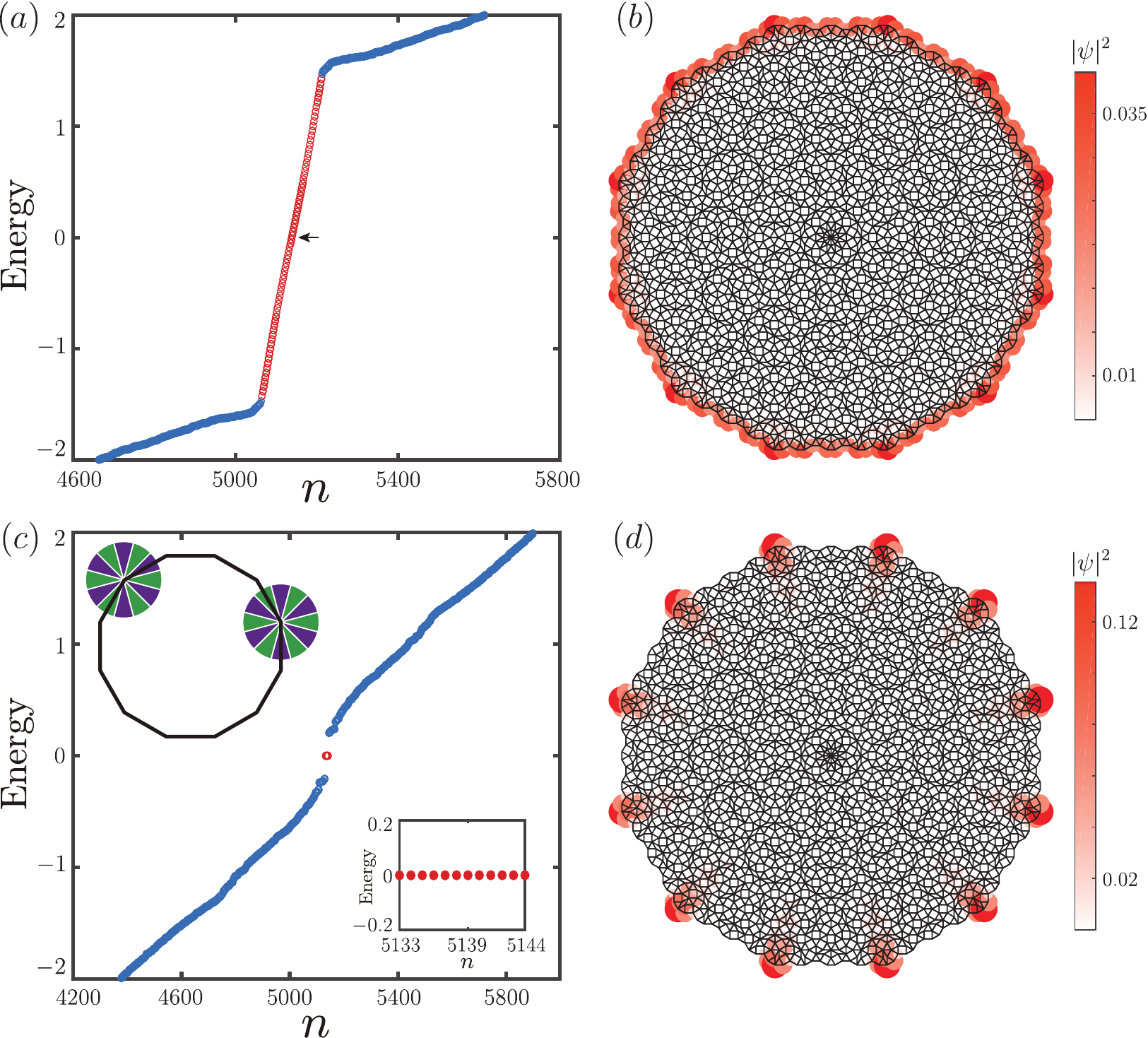} \caption{(a) Energy spectrum of the first-order topological insulator Hamiltonian $H_{\text{0}}$ on a quasicrystal dodecagon vs the eigenvalue index $n$. Red circles mark all the edge states. (b) The probability density of doubly degenerate eigenstates near zero energy marked by the black arrow in (a). (c) Energy spectrum of the HOTI Hamiltonian $H$ versus the eigenvalue index $n$ for the TRS breaking mass $g/t_{2}=2$. The inset at top shows the color circle of the effective edge mass. The green and violet regions denote the two regions of the edge orientation with opposite sign of the effective edge mass. The inset at lower right shows the enlarged section of 12 zero-energy modes marked by the red dots. (d) The probability density of zero-energy modes in (c). The color map shows the values of the probability density. We take the model parameters $t_{1}/t_{2}=2$, $M/t_{2}=1$, and lattice site number $N=2569$.}%
\label{fig2}
\end{figure}

To study the topological edge states, we directly diagonalize the Hamiltonian (\ref{H0}) on a dodecagon geometry under the open boundary condition. The energy spectrum versus the eigenvalue index $n$ is illustrated in Fig.~\ref{fig2}(a), and the red circles depicts the eigenvalues of edge states. Note that all the edge states are doubly degenerate counterpropagating modes due to TRS. Figure~\ref{fig2}(b) shows the spatial probability density of a pair of doubly degenerate edge states around zero energy indicated by a black arrow in Fig.~\ref{fig2}(a). We can see that the edge states are well localized at the sample edge. Recently, 2D topological insulator phase has been also proposed on the fivefold Penrose tiling QL \cite{PhysRevLett.121.126401,PhysRevB.98.125130} and the eightfold AB tiling QL \cite{PhysRevB.100.085119,PhysRevLett.124.036803}.

\emph{\color{blue}Corner states on the dodecagonal QL.}---To realize the 2D quasicrystalline HOTI, we gap out 1D edge state in the first-order topological insulator by introducing a symmetry-breaking mass. Recently, this approach has been used in realizing HOTIs in periodic crystals~\cite{Langbehn2017PRL,Song2017PRL,Schindler2018SA} and aperiodic quasicrystals~\cite{PhysRevLett.124.036803, Varjas2019PRL}. Similarly, we introduce a TRS breaking mass term to the dodecagonal QL we designed
\begin{equation}
H_{m}=g\sum_{j\not=k}\frac{u\left(d_{jk}\right)}{2}\cos\left(\eta\phi_{jk}\right)c_{j}^{\dag }\sigma_{x}\tau_{x}c_{k},
\label{Hm}
\end{equation}
where $g$ denote the magnitude of the mass term, and the parameter $\eta$ in cosine function represents the spatially varying period of the mass term. Now, the total Hamiltonian of HOTI on the QL is $H=H_{0}+H_{m}$. For the dodecagonal QL, we choose $\eta=6$ to match the underlying symmetry. The case of $\eta=2$ and $\eta=4$ can give rise to two distinct types of HOTIs associated with the fourfold and eightfold rotational symmetries in the AB tiling quasicrystal, which has been discussed in Ref. \cite{PhysRevLett.124.036803}. When $\eta=6$, the non-zero mass term $H_m$ in the dodecagonal quasicrystal breaks the rotational symmetry $C_{12}$ and mirror symmetry $m_z$ in addition to TRS.

After turning on the mass term $H_m$, we numerically diagonalize the HOTI Hamiltonian $H$ on a dodecagon-shaped Stampfli-tiling QL.
We plot the energy spectrum versus the eigenvalue index $n$ in Fig.~\ref{fig2}(c), and found that $H_m$ opens an energy gap in the edge spectrum. Meanwhile, twelve zero-energy modes appear in the edge energy gap. The spatial probability density of zero-energy modes is shown in Fig.~\ref{fig2}(d). We can see that the zero-energy modes are symmetrically localized at the twelve corners of the regular quasicrystal dodecagon. The twelvefold symmetric zero-energy corner modes are a hallmark feature of the 2D quasicrystalline HOTI on the dodecagonal QL.

The corner states are protected by the combination of rotational symmetry $C_{12}$ and mirror symmetry $m_z$. PHS pins the corner states to zero energy.  More details of symmetry analysis for the system are given in Ref.~\cite{SM}. This kind of symmetry-protected corner states in the dodecagonal quasicrystal is quite distinct from those in crystalline systems, because the rotational symmetry $C_{12}$ is a forbidden rotational symmetry in a crystal. 

\emph{\color{blue}Stability of corner states.}---
Here, we use symmetry-breaking perturbations to examine the robustness of the zero-energy corner modes. As mentioned above, the mass term $H_{m}$ required by the quasicrystalline HOTI breaks TRS, the mirror symmetry $m_{z}$, and the rotational symmetry $C_{12}$. However, the combined symmetries $C_{12}m_z$ and $C_{12}T$ are preserved. In the following, we use on-site potential perturbations in the calculations, which can be written as
\begin{equation}
\Delta H^{pq}=U\sum_{j}c_{j}^{\dag }\sigma _{p} \tau_{q}c_{j},
\end{equation}
where $U$ is the potential strength, $p$ and $q=0,x,y,z$ denote the $2\times2$ identity matrix and the three components of the Pauli matrices, respectively. There are sixteen perturbations in total, but four of them ($\Delta H^{0y}$, $\Delta H^{0z}$, $\Delta H^{xx}$ and $\Delta H^{zx}$) can be absorbed into the Hamiltonian, which are neglected in our calculations.

\begin{figure}[t]
	\includegraphics[width=8cm]{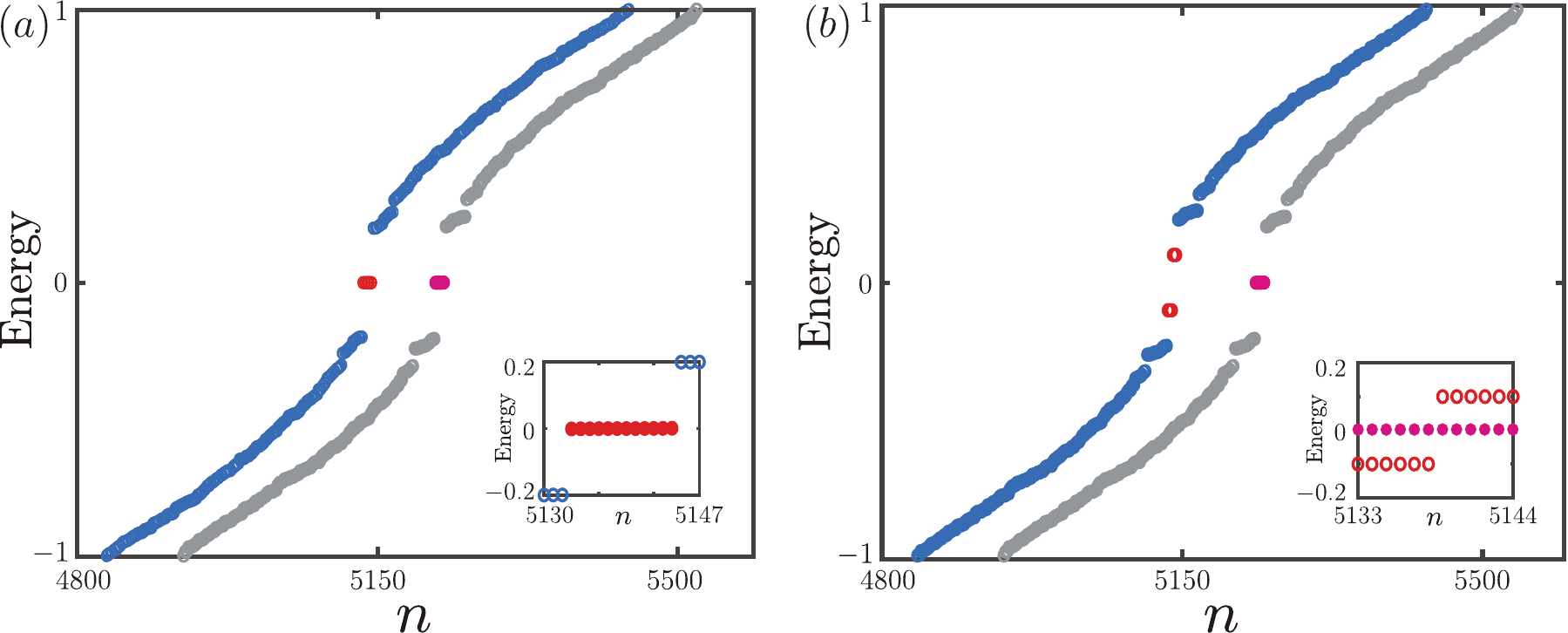} \caption{Energy spectra of (a) $H+\Delta H^{zz}$ and (b) $H+\Delta H^{yx}$ on the dodecagonal QL versus the eigenvalue index $n$. For comparison, we also plot the energy spectrum (shown in gray circles and pink dots) without any perturbations. The insets show the enlarged section of the corner state region. We take the parameters $t_{1}/t_{2}=2$, $M/t_{2}=1$, $g/t_{2}=2$, $U/t_{2}=0.1$, and $N=2569$.}%
\label{fig3}
\end{figure}

The zero-energy corner states in the dodecagonal quasicrystal remain stable under weak perturbations as long as PHS preserves or the effective chiral symmetry \cite{SM}, which is the product of chiral symmetry and $m_z$, exists. In the main text, we demonstrate the results of two kinds of perturbations, which are $\Delta H^{zz}=U\sum_{j}c_{j}^{\dag }\sigma _{z} \tau_{z}c_{j}$ and $\Delta H^{yx}=U\sum_{j}c_{j}^{\dag }\sigma _{y} \tau_{x}c_{j}$. The energy spectra of the quasicrystalline HOTI Hamiltonian in the presence of perturbations $\Delta H^{zz}$ and $\Delta H^{yx}$ are depicted in Figs.~\ref{fig3}(a) and \ref{fig3}(b), respectively. Note that $\Delta H^{yx}$ breaks PHS, while $\Delta H^{zz}$ doesn't. Therefore, the twelve zero-energy corner states are robust under the perturbation $\Delta H^{zz}$, as shown in Fig.~\ref{fig3}(a). In contrast, in Fig.~\ref{fig3}(b) we can see that the zero-energy corner states are gapped out by $\Delta H^{yx}$. We also summarize the results of other kinds of perturbations including two with the effective chiral symmetry in Ref. \cite{SM}.

It is necessary to point out that the combined spatial symmetry $C_{12}m_z$ protects the zero-energy corner modes against boundary perturbations or deformations that are compatible with $C_{12}m_z$. For instance, attaching every edge of the sample with a 1D topological insulator with zero-energy end states cannot remove the corner states. The role of spatial symmetries in topological protection of corner states has been discussed in Ref. \cite{Varjas2019PRL}.

\emph{\color{blue}Effective edge theory of corner states.}---
The physical explanation of the zero-energy corner states in HOTIs can be given by the Jackiw-Rebbi mechanism \cite{JRPRD1976}. In this mechanism, a topological zero-energy mode appears when a mass domain wall forms. In the present case, the mass term $H_m$, relying on the polar angle of the bond $\phi_{jk}$,  can result in an effective edge mass domain structure. However, it is unlikely to derive an explicitly analytic expression of the effective mass for the edge states on the dodecagonal QL, owing to the lack of translational symmetry. However, as a rough approximation, we can treat the sides of a quasicrystal polygon as a long ``bond" and the sign of the effective mass for the edge state depends on the polar angle of the sides $\theta_{\text{edge}}$ \cite{PhysRevLett.124.036803}. For a given side of quasicrystal polygons, the effective edge mass on the side is determined by the factor $\cos\left(6\theta_{jk}\right)$, which controls the sign of effective mass by varying $\theta_{\text{edge}}$.

The green and violet in the top inset of Fig.~\ref{fig2}(c) define different regions with opposite signs of the effective edge mass. The green region is determined by $\theta_{\text{edge}}\in ( -\frac{\pi }{12}+\frac{n\pi }{3},\frac{\pi }{12}+\frac{n\pi }{3})$, while the violet region is $\theta_{\text{edge}}\in ( \frac{\pi }{12}+\frac{n\pi }{3},\frac{3\pi }{12}+\frac{n\pi }{3})$, where $n=0,1,2,3,4,5$. According to $\theta_{\text{edge}}$, we can tell whether there is a zero-energy mode at a boundary corner. For the Stampfli-tiling QL with regular dodecagon boundary, all the adjacent sides lie in two different regions, as shown in the top inset of Fig.~\ref{fig2}(c), thus an effective mass domain wall occurs at all the corners of the regular dodecagon, resulting in twelve ingap zero-energy modes.

\begin{figure}[t]
	\includegraphics[width=8cm]{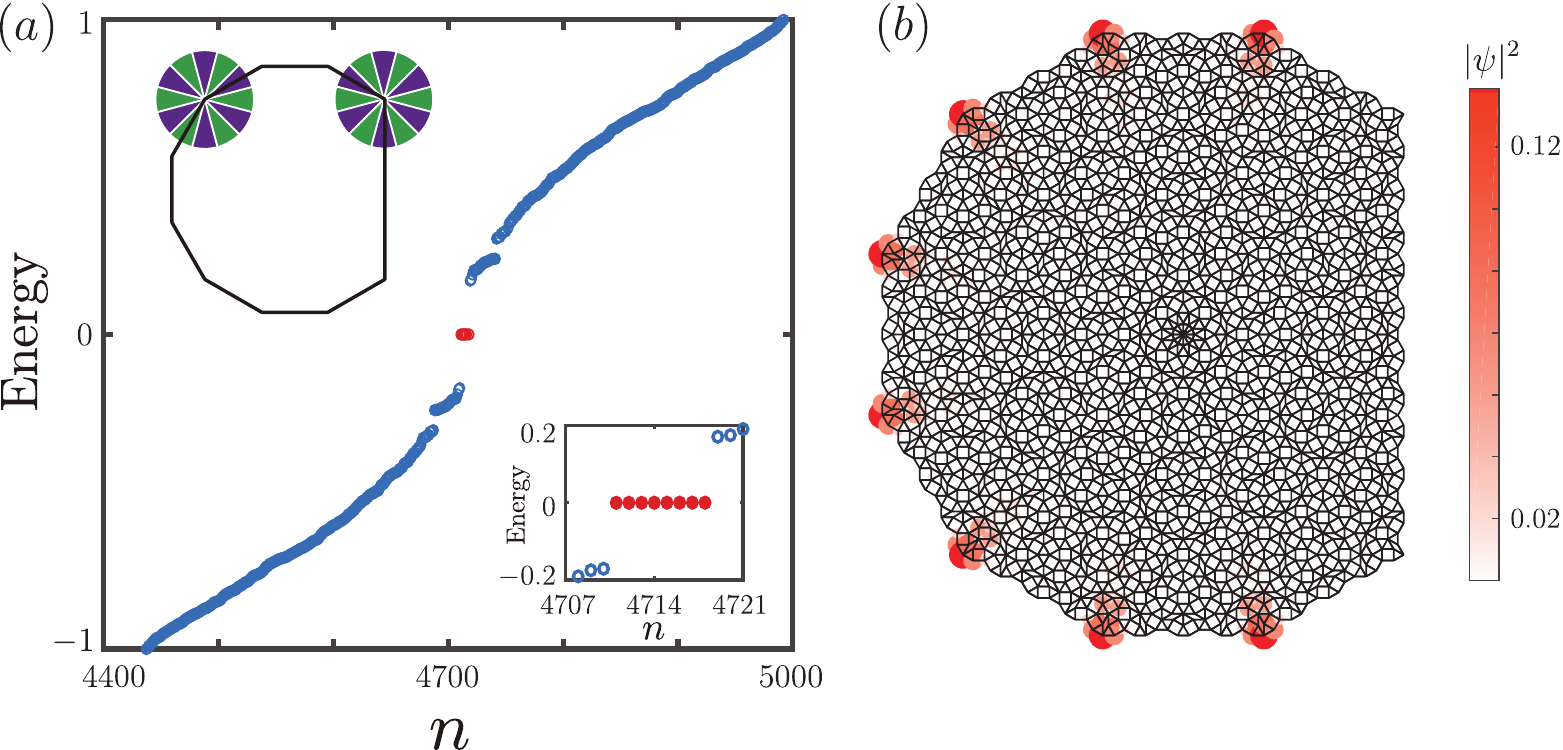} \caption{(a) Energy spectrum of $H$ on a quasicrystal polygon with ten vertices versus the eigenvalue index $n$. The inset at top shows the color circle of the effective edge mass. The green and violet regions denote the two regions of the edge orientation with opposite sign of the effective edge mass. The  bottom inset shows the enlarged section of eight zero-energy modes marked by the red dots. (b) The probability density of eight zero-energy modes in (a). The color map shows the values of the probability density. We take the parameters $t_{1}/t_{2}=2$, $M/t_{2}=1$, and $g/t_{2}=2$.}%
\label{fig4}
\end{figure}

To further illustrate the validity of the effective edge theory, now we consider the dodecagonal QL with a different boundary shape in Fig.~\ref{fig4}. This geometric structure is obtained by cutting off four vertices/corners of a quasicrystal regular dodecagon. The energy spectrum of the Hamiltonian $H$ for this geometric structure is shown in Fig.~\ref{fig4}(a). We found that eight zero-energy modes appear in the energy gap of the edge spectrum. Figure~\ref{fig4}(b) shows the spatial probability density of the eight zero-energy modes, we can see that they are localized at the eight vertices of the quasicrystal regular dodecagon. The corner states are absent at the two rightmost vertices of the quasicrystal polygon. We can also use the effective edge theory to explain why there are only eight zero-energy corner states in a quasicrystal polygon with ten vertices. For the two rightmost corners, all adjacent sides fall into the regions with the same effective mass sign, therefore, no mass domain wall forms at these two corners and no zero-energy bound mode appears. Note that, for this quasicrystal polygon, we cannot define a global rotational symmetry, thus the corner states lack the protection available for the previous twelvefold symmetric corner states. Moreover, we also show the results on the dodecagonal QL under different boundary shapes in Ref. \cite{SM}. We found that the numerical results are in good agreement with the effective edge theory based on the Jackiw-Rebbi mechanism.

\emph{\color{blue}Conclusion and Discussion.}---
In this work, we realize the 2D quasicrystalline HOTI in the Stampfli-tiling dodecagonal quasicrystal. The quasicrystalline HOTI is obtained by introducing an additional mass term to the first-order topological insulator. For a finite-sized quasicrystal dodecagon, the quasicrystalline HOTI hosts twelve zero-energy corner states which are protected by the combination of the twelvefold rotational symmetry and mirror symmetry, as well as PHS. The corner states are robust against any symmetry-preserving perturbations due to the topological protection provided by the symmetries.

The quasicrystalline HOTIs protected by forbidden symmetries in crystals are beyond the present topological classification based on crystalline symmetries, thus they cannot be properly characterized by previously known topological invariants. The 2D quasicrystalline HOTI in the AB tilling octagonal quasicrystal is characterized by a $\mathbb{Z}_2$ invariant defined in the hyperspace \cite{PhysRevLett.124.036803,Varjas2019PRL}. However, it is not easy to extend this method to other quasicrystalline systems including the present Stampfli-tiling QL. A topological invariant to the quasicrystalline HOTI in the dodecagonal quasicrystal will be investigated in our future work.

\emph{\color{blue}Acknowledgments.}---
This work was financially supported by the NSFC (Grant Nos. 11704106, 12074108, and 12074107). D.-H.X. also acknowledges the financial support of the Chutian Scholars Program in Hubei Province. R.C. was supported by the Project funded by China Postdoctoral Science Foundation (Grant No. 2019M661678).

\emph{\color{blue}Note added.}---
Recently, we became aware of a complementary study \cite{PhysRevResearch.2.033071}, which addresses similar problems from a different perspective.

\bibliographystyle{apsrev4-1-etal-title_6authors}
\bibliography{bibfile}

\pagebreak
\widetext
\clearpage

\setcounter{equation}{0}
\setcounter{figure}{0}
\setcounter{table}{0}
\makeatletter
\renewcommand{\tablename}{TAB.}
\renewcommand{\theequation}{S\arabic{equation}}
\renewcommand{\thetable}{S\arabic{table}}
\renewcommand{\thefigure}{S\arabic{figure}}
\renewcommand{\bibnumfmt}[1]{[S#1]}
\renewcommand{\citenumfont}[1]{S#1}

\begin{center}
\textbf{\large Supplemental Material to: ``Higher-order topological insulator in a dodecagonal quasicrystal''}
\end{center}

In this Supplemental Material, we first describe the algorithm of the process used to construct the Stampfli-tiling dodecagonal quasicrystals in Sec.~\ref{I}. Then, in Sec.~\ref{II} we present a symmetry analysis of the quasicrystalline system and the stability of zero-energy corner states in the quasicrystalline higher-order topological insulator (HOTI). At last, we show the corner states on the Stampfli-tiling quasicrystalline lattice (QL) with different boundary shapes Sec.~\ref{III}.

\section{Dodecagonal Tiling construction}
\label{I}
\begin{figure}[hptb]
	\includegraphics[width=14cm]{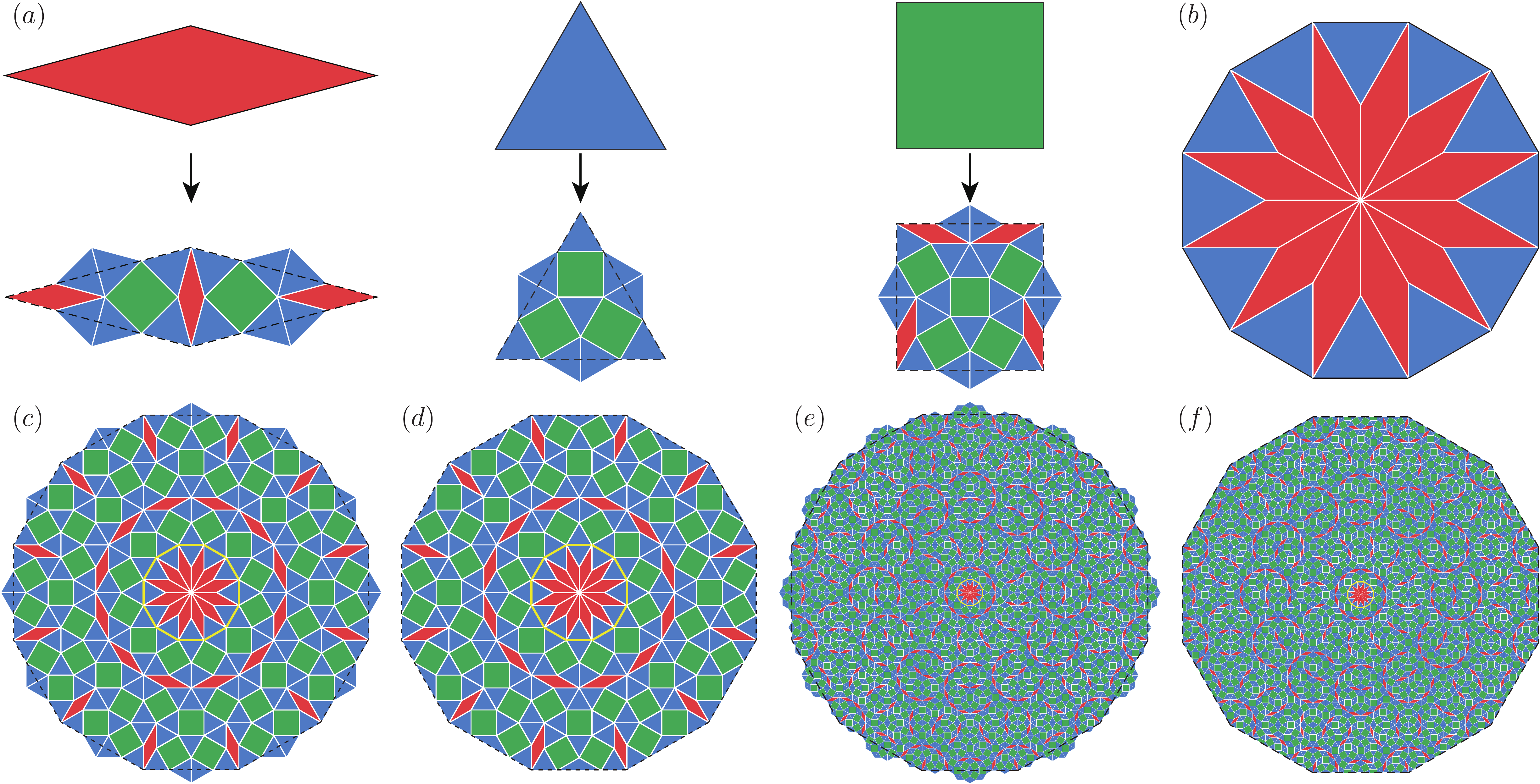} \caption{Obtaining the dodecagonal tiling through inflation method. (a) The inflation rules. (b) The initial regular dodecagonal tiling. (c) and (e) The dodecagonal tiling obtained after $d=1$ and $d=2$ applications of the rules. (d) and (f) The dodecagonal quasicrystals after removing the redundant vertices.}%
\label{figS1}
\end{figure}

Dodecagonal quasicrystals possess local twelvefold rotational symmetry, and it was first proposed by P. Stampfli \cite{stampfli1986dodecagonal2}. The Stampfli tiling consists of three primitive tiles: a square, a regular triangle, and a rhombus with the small angle equal to ${30}^{\ensuremath{\circ}}$. The large scale structure of the dodecagonal tiling can be formed by the inflation method \cite{baake2013aperiodic2} of the primitive tiles. As shown in Fig.~\ref{figS1}(a), a rhombus is divided into 3 small rhombuses, 12 small regular triangles, and 2 small squares; a regular triangle is divided into 10 small regular triangles and 3 small squares; a square is transformed into 4 small rhombuses, 20 small regular triangles, and 5 small squares.

To obtain a dodecagonal quasicrystal, we start with a regular dodecagon (shown in Fig.~\ref{figS1}(b)), and then apply the inflation method as mentioned above to the regular dodecagon $d$ times. The twelvefold rotational symmetry is maintained before and after the transformation. After the inflation, a large scale dodecagonal quasicrystal is obtained. The dodecagonal quasicrystal obtained by applying $d=1$ and $d=2$ times the inflation to the regular dodecagon in Fig.~\ref{figS1}(b) are illustrated in Figs.~\ref{figS1}(c) and \ref{figS1}(e), where the number of vertices are respectively $217$ and $2713$. Furthermore, to get a dodecagonal QL with a perfect regular dodecagon boundary, we remove all the vertices outside the original dodecagonal edges (as shown in the black dotted lines in the Fig.~\ref{figS1}(c) and (e)). Figures~\ref{figS1}(d) and (f) are the desirable dodecagonal quasicrystals after removing the redundant vertices, where the number of vertices are respectively $205$ and $2569$.

\section{Symmetry analysis and perturbations to corners states}
\label{II}
In this Section, we present a symmetry analysis of the system, and then adopt some symmetry-breaking perturbation terms to examine the robustness of the zero-energy corner states. In the symmetry analysis, the following symmetries are involved: particle-hole symmetry (PHS) $P$, time-reversal symmetry (TRS) $T$, chiral symmetry $S$, mirror symmetries $m_{i}$ with $i=x,y,z$, as well as the rotational symmetry $C_{12}$. Table~\ref{tabS1} illustrates the symmetries of the first-order topological insulator Hamiltonian $H_{0}$ [Eq.~(1) in the main text] on the dodecagonal QL. It is found that this Hamiltonian preserves particle-hole, time-reversal, and the chiral symmetries, and thus belongs to the class DIII of the ten-fold Altland-Zirnbauer classification \cite{AZ1997PRB2,Schnyder2008PRB2,Schnyder2009AIP2,Ryu2010NJP2,Chiu2016RMP2}. In addition, the first-order topological insulator in a finite-sized Stampfli-tiling QL with dodecagon boundary also preserves three mirror symmetries and a twelvefold rotational symmetry as shown in the Tab.~\ref{tabS1}.

\begin{table}[htpb]
    \centering
\begin{tabular}{|c|c|c|c|}
\hline
&&$g=0$&$g\not=0$\cr
\hline
$P=\sigma_{z}\tau_{x}\mathcal{I}K$  &$PHP^{-1}=-H$                     &\checkmark &\checkmark             \\
\hline
$T=i \sigma_{y}\tau_{0}\mathcal{I}K$                        &$THT^{-1}=H$&\checkmark &$\times$        \\
\hline
$S=PT$                        &$SHS^{-1}=-H$ &\checkmark      &$\times$            \\
\hline
$m_{x}=\sigma_{x}\tau_{0}\mathcal{M}_{x}$&$m_{x}Hm_{x}^{-1}=H$&\checkmark &\checkmark       \\
\hline
$m_{y}=\sigma_{y}\tau_{z}\mathcal{M}_{y}$&$m_{y}Hm_{y}^{-1}=H$&\checkmark &\checkmark          \\
\hline
$m_{z}= \sigma_{z}\tau_{0}\mathcal{I}$&$m_{z}Hm_{z}^{-1}=H$&\checkmark &$\times$    \\
\hline
$C_{12}=e^{-i\frac{\pi}{12}\sigma_{z}\tau_{z}}\mathcal{R}_{12}$&$C_{12}HC_{12}^{-1}=H$&\checkmark       &   $\times$            \\
\hline
$C_{12}T$&$C_{12}TH(C_{12}T)^{-1}=H$&\checkmark      &\checkmark            \\
\hline
$C_{12}m_{x}$&$C_{12}m_{x}H(C_{12}m_{x})^{-1}=H$&\checkmark      &$\times$              \\
\hline
$C_{12}m_{y}$&$C_{12}m_{y}H(C_{12}m_{y})^{-1}=H$&\checkmark      &$\times$                \\
\hline
$C_{12}m_{z}$&$C_{12}m_{z}H(C_{12}m_{z})^{-1}=H$&\checkmark      &\checkmark                 \\
\hline
\end{tabular}
    \caption{Symmetries of the HOTI Hamiltonian $H$ on a Stampfli-tiling QL with dodecagon boundary without ($g=0$) and with ($g\neq0$) the mass term $H_m$. $K$ is the complex conjugate operator, and $\mathcal{I}$ is the $N\times N$ unit matrix with the lattice number $N$. $\mathcal{M}_{x,y}$ are orthogonal matrices permuting the sites of the tiling to flip the whole system vertically and horizontally. $\mathcal{R}_{12}$ is an orthogonal matrix permuting the sites of the tiling to rotate the whole system by an angle of $\pi/6$. Check mark indicates that the symmetry in this case is preserved, and a cross mark means the symmetry is absent.}
    \label{tabS1}
\end{table}

\begin{table}[htpb]
    \centering
\begin{tabular}{|c|c|c|c|c|c|c|c|c|c|c|c|c|c|}
\hline
\multirow{2}{*}{}&\multirow{2}{*}{$U=0$}&\multicolumn{12}{c|}{$U\not=0$}
\cr\cline{3-14}&&$00$&$0x$&$x0$&$xy$&$xz$&$y0$&$yx$&$yy$&$yz$&$z0$&$zy$&$zz$\cr
\hline
$P$                        &\checkmark &$\times$      &$\times$            &\checkmark  &\checkmark &$\times$
                                                     &$\times$      &$\times$      &$\times$     &\checkmark &$\times$            &$\times$      &\checkmark \\
\hline
$T$                        &$\times$      &$\times$      &$\times$           &$\times$       &$\times$      &$\times$
                                                     &$\times$      &$\times$      &$\times$      &$\times$     &$\times$             &$\times$       &$\times$ \\
\hline
$S$                        &$\times$      &$\times$      &$\times$                  &$\times$            &$\times$      &$\times$
                                                     &$\times$      &$\times$      &$\times$      &$\times$      &$\times$            &$\times$      &$\times$ \\
\hline
$m_{x}$               &\checkmark &\checkmark &\checkmark   &\checkmark  &\checkmark &\checkmark
                                                     &$\times$      &$\times$      &$\times$      &$\times$      &$\times$             &$\times$       &$\times$ \\
\hline
$m_{y}$               &\checkmark &\checkmark &$\times$             &$\times$        &\checkmark &$\times$
                                                     &\checkmark &$\times$      &$\times$      &\checkmark &$\times$        &\checkmark &$\times$ \\
\hline
$m_{z}$               &$\times$      &$\times$      &$\times$                   &$\times$             &$\times$      &$\times$
                                                    &$\times$      &$\times$      &$\times$       &$\times$      &$\times$              &$\times$      &$\times$ \\
\hline
$C_{12}$             &$\times$      &$\times$      &$\times$                   &$\times$       &$\times$            &$\times$
                                                    &$\times$      &$\times$      &$\times$      &$\times$       &$\times$             &$\times$     &$\times$ \\
\hline
$C_{12}T$          &\checkmark &\checkmark &$\times$             &$\times$              &\checkmark &$\times$
                                                    &$\times$      &$\times$      &\checkmark &$\times$      &$\times$              &$\times$       &$\times$ \\
\hline
$C_{12}m_{x}$ &$\times$      &$\times$      &$\times$             &$\times$              &$\times$      &$\times$
                                                   &$\times$      &$\times$       &$\times$      &$\times$      &$\times$               &$\times$      &$\times$ \\
\hline
$C_{12}m_{y}$ &$\times$      &$\times$      &$\times$            &$\times$              &$\times$      &$\times$
                                                   &$\times$      &$\times$       &$\times$      &$\times$       &$\times$              &$\times$     &$\times$ \\
\hline
$C_{12}m_{z}$ &\checkmark &\checkmark &$\times$       &$\times$              &$\times$      &$\times$
                                                   &$\times$      &$\times$      &$\times$       &$\times$      &\checkmark          &$\times$      &\checkmark \\
\hline
\end{tabular}
    \caption{Symmetry analysis of the quasicrystalline HOTI without ($U=0$) and with ($U\neq0$) the perturbation terms $\Delta H^{pq}$.}
    \label{tabS2}
\end{table}

The symmetry-breaking mass term $H_{m}$ [Eq.~(4) in the main text] breaks TRS, $m_z$ and $C_{12}$, but preserves PHS, $m_{x}$ and $m_{y}$, as well as the combined symmetries $C_{12}T$ and $C_{12}m_{z}$.

\begin{figure}[t]
	\includegraphics[width=16cm]{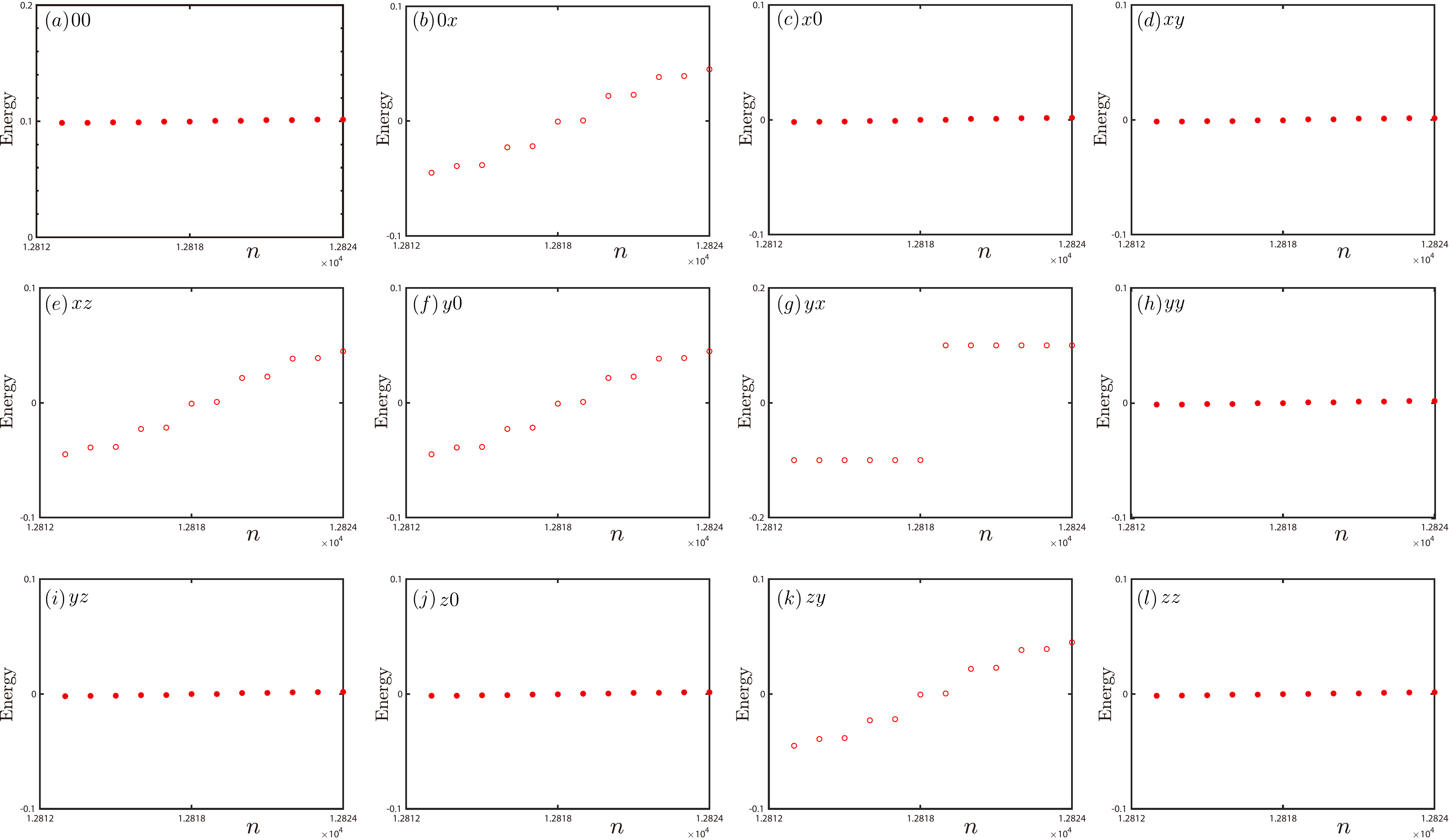} \caption{(a)-(l) Energy spectra of $H+\Delta H^{pq}$ in the Stampfli-tiling quasicrystal dodecagon versus the eigenvalue index $n$. We take the parameters $t_{1}/t_{2}=2$, $M/t_{2}=1$, $g/t_{2}=3$, the on-site potential strength $U/t_{2}=0.1$, and the lattice number $N=6409$.}%
\label{figS2}
\end{figure}

The twelvefold symmetric zero-energy corner states in the Stampfli-tiling quasicrystal dodecagon are protected by PHS and the combined symmetry $C_{12}m_z$. As stated in the main text, we employ on-site potential perturbations to test the stability of the corner states. Actually, there are in total 16 kinds of on-site potentials $\Delta H^{pq}=U\sum_{j}c_{j}^{\dag }\sigma _{p} \tau_{q}c_{j}$ with $p,q=0,x,y,z$. 12 kinds of them are listed in Tab.~\ref{tabS2}, and the rest four kinds ($\Delta H^{0y}$, $\Delta H^{0z}$, $\Delta H^{xx}$ and $\Delta H^{zx}$) are neglected due to they can be absorbed into the Hamiltonian. The symmetries of the HOTI under perturbations $\Delta H^{pq}$ are also illustrated in Tab.~\ref{tabS2}. We plot the energy spectra of the HOTI Hamiltonian with different perturbations on the dodecagonal QL in Fig.~\ref{figS2}.

We note that the corner states are pinned at zero energy due to PHS and the effective chiral symmetry $S_{\text{eff}}=\sigma_{y}\tau_{x}\mathcal{I}$ which is just the product of chiral symmetry and $m_z$. As shown in Fig.~\ref{figS2}, the zero-energy corner states remain stable in the presence of six kinds of on-site potential perturbations, which are $\Delta H^{x0}$ [Fig.~\ref{figS2}(c)], $\Delta H^{xy}$ [Fig.~\ref{figS2}(d)], $\Delta H^{yy}$ [Fig.~\ref{figS2}(h)], $\Delta H^{yz}$ [Fig.~\ref{figS2}(i)], $\Delta H^{z0}$ [Fig.~\ref{figS2}(j)] and $\Delta H^{zz}$ [Fig.~\ref{figS2}(l)]. In Figs.~\ref{figS2}(c), (d), (i), and (l), the zero-energy in-gap states under perturbations remain stable as PHS exists, while in Figs.~\ref{figS2}(h) and (j), the corner states subjected to perturbations are still at zero energy as the effective chiral symmetry is preserved.

\section{Quasicrystalline HOTI on the QL with other boundaries}
\label{III}

\begin{figure}[t]
	\includegraphics[width=8cm]{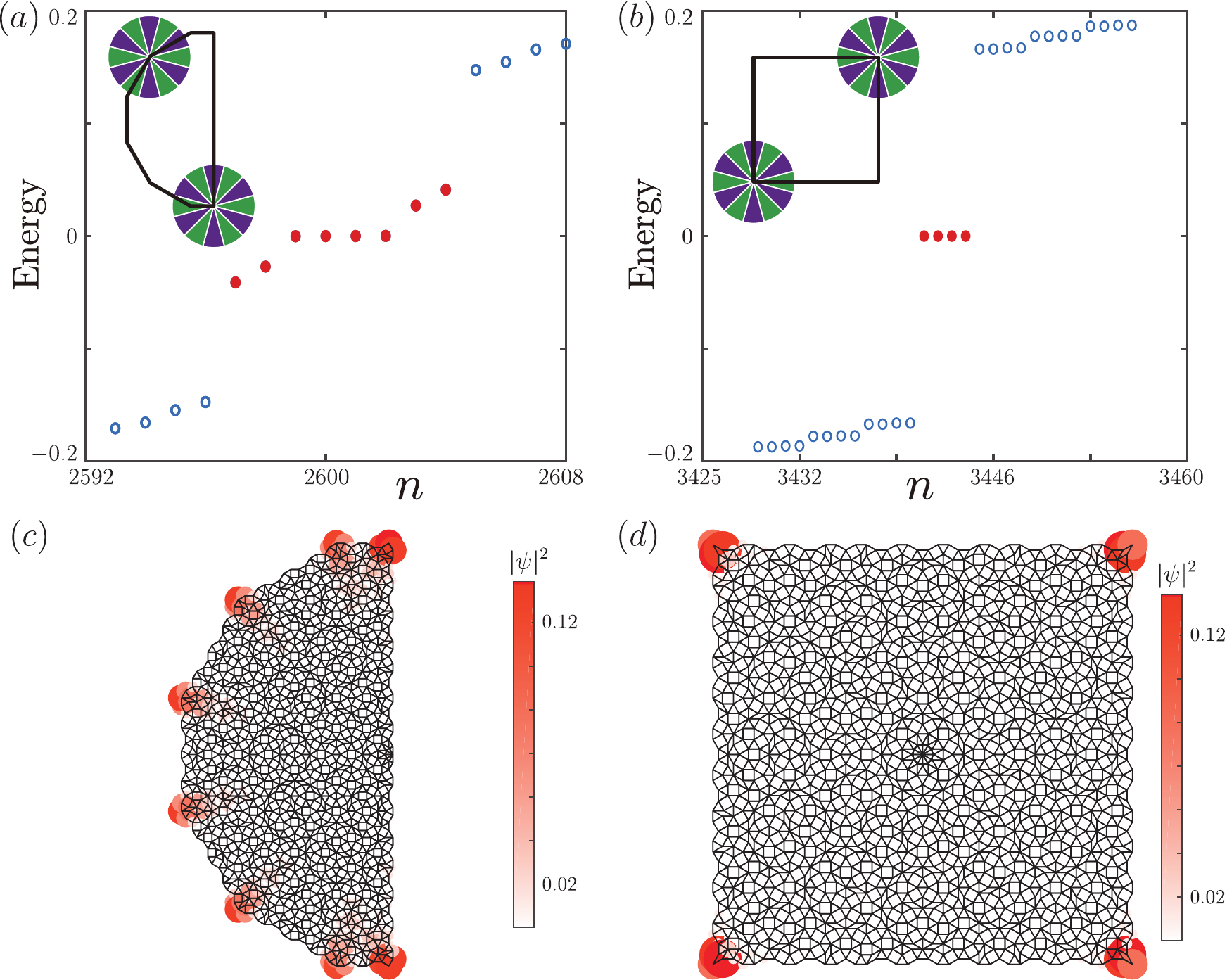} \caption{(a) and (b) Energy spectra of the quasicrystalline HOTI versus the eigenvalue index $n$ on the dodecagonal QL with different boundaries.The inset at top shows the color circle of the effective edge mass.The green and violet regions denote the two regions of the edge orientation with opposite sign of the effective edge mass. (c) and (d) The probability density of in-gap states marked by the red dots in (a) and (b), respectively. The color map shows the values of the probability density. We take the parameters $t_{1}/t_{2}=2$, $M/t_{2}=1$, and $g/t_{2}=2$.}%
	\label{figS3}
\end{figure}

In this section, we show the energy spectrum and probability density of corner states for the Stampfli-tiling QL with different boundary shapes. We can also use the effective edge theory based on the Jackiw-Rebbi mechanism to determine the location of the corner states. Note that there are eight corner states in Figs.~\ref{figS3}(a) and \ref{figS3}(c). However, due to the size effect, four of them are gapped out. For the squared boundary, there are four zero-energy corner states as shown in Figs. \ref{figS3}(b) and \ref{figS3}(d). We can control the number of corner sates by designing the boundary shapes.

\end{document}